\documentclass[prl,twocolumn,showpacs,amsmath,superscriptaddress,psfig,amssymb]{revtex4}

\usepackage{graphicx}
\usepackage{dcolumn}
\usepackage{bm}

\begin{document}


\title{Weak anisotropy of the superconducting upper critical field in Fe$_{1.11}$Te$_{0.6}$Se$_{0.4}$ single crystals}

\author{M. H. Fang}
\affiliation {Department of Physics, Zhejiang University, Hangzhou, Zhejiang 310027, China}

\author{J. H. Yang}
\affiliation {Department of Physics, Zhejiang University, Hangzhou, Zhejiang 310027, China}

\author{ F. F. Balakirev}
\affiliation {NHMFL, Los Alamos National Laboratory, MS E536, Los Alamos, NM 87545, USA}

\author{Y. Kohama}
\affiliation {NHMFL, Los Alamos National Laboratory, MS E536, Los Alamos, NM 87545, USA}

\author{J. Singleton}
\affiliation {NHMFL, Los Alamos National Laboratory, MS E536, Los Alamos, NM 87545, USA}

\author{B. Qian}
\affiliation { Department of Physics, Tulane University, New Orleans, LA70118, USA}

\author{Z. Q. Mao}
\affiliation { Department of Physics, Tulane University, New Orleans, LA70118, USA}

\author{H. D. Wang}
\affiliation {Department of Physics, Zhejiang University, Hangzhou, Zhejiang 310027, China}

\author{H. Q. Yuan}
\email{hqyuan@zju.edu.cn}
\affiliation {Department of Physics, Zhejiang University, Hangzhou, Zhejiang 310027, China}

\date{\today}

\begin{abstract}
\noindent We have determined the resistive upper critical field $H_{\rm c2}$ for single crystals of the superconductor Fe$_{1.11}$Te$_{0.6}$Se$_{0.4}$ using pulsed magnetic fields of up to 60T. A rather high zero-temperature upper critical field of $\mu_0 H_{c2}(0) \approx 47$T is obtained, in spite of the relatively low superconducting transition temperature ($T_{\rm c} \approx 14$K). Moreover, $H_{\rm c2}$ follows an unusual temperature dependence, becoming almost independent of the magnetic field orientation as the temperature $T \rightarrow 0$. We suggest that the isotropic superconductivity in Fe$_{1.11}$Te$_{0.6}$Se$_{0.4}$ is a consequence of its three-dimensional Fermi-surface topology. An analogous result was obtained for (Ba,K)Fe$_2$As$_2$, indicating that all layered iron-based superconductors exhibit generic behavior that is significantly different from that of the ``high-$T_{\rm c}$'' cuprates.
\end{abstract}

\maketitle

The discovery of superconductivity in the iron pnictides LnFeAs(O,F) (where Ln can be La, Ce, Pr, Nd, Sm or Gd) \cite{Kamihara 2008, Chen 2008, Chen 2008-2, Wen 2008, Ren 2008} with transition temperatures $T_{\rm c}$ as high as 55~K has been responsible for something of a resurrection in the study of high temperature superconductivity. Beside the LnFeAs(O,F) series (the so-called ``1111s''), other families of the iron-based superconductors have been found, including the ``122'' materials possessing the ThCr$_2$Si$_2$ structure ({\it e.g.}, hole- or electron- doped BaFe$_{2}$As$_{2}$) \cite{Rotter 2008, Sefat 2008}, the ``111-type'' LiFeAs family \cite{Wang 2008, Pitcher 2008} and the ``11-type'' iron chalcogenides with an $\alpha-$PbO structure ({\it e.g.}, Fe$_{1+x}$(Se,Te) \cite{Hsu 2008, Fang 2008}). All of these compounds share a common structural feature, {\it i.e.}, square planar sheets of Fe, coordinated tetrahedrally by pnictogens or chalcogens. The relatively high superconducting transition temperatures and layered crystal structures of the Fe-based superconductors initially suggested strong analogies with the cuprates. However, in this letter we report pulsed-field magnetoresistance measurements for single crystals of Fe$_{1.11}$Te$_{0.6}$Se$_{0.4}$ that show that its upper critical field attains a value of 47T as temperature $T \rightarrow 0$ that is almost independent of field direction. This suggests that the electronic properties of Fe$_{1+x}$(Te,Se) superconductors are rather isotropic ({\it i.e.}, three dimensional), in complete contrast to those of the quasi-two-dimensional cuprates. A similar effect was found in (Ba,K)Fe$_2$As$_2$ \cite{Yuan 2009} and other 122-type systems \cite{Baily 2009, Altarawneh 2008, Kano 2009}, indicating that this may be a general feature of {\it all} iron pnictides.

Large single crystals of Fe$_{1.11}$Te$_{0.6}$Se$_{0.4}$ were grown by a self-flux method. The starting composition was Fe(Te$_{0.6}$Se$_{0.4}$)$_{0.85}$. The mixtures of Fe, and (Te,Se) were ground thoroughly and sealed in an evacuated quartz tube. The tube was heated to $920^{\circ}$C and cooled slowly to grow large single crystals. The crystals obtained were checked by X-ray diffraction (XRD); their composition was analyzed using a scanning electron microscope (Hitachi S3400) equipped with an Energy Dispersive X-Ray Spectrometer (EDXS). Longitudinal resistivity was measured using a typical four-contact method in pulsed fields of up to 60T at the National High Magnetic Field Laboratory, Los Alamos \cite{Yuan 2009}. In order to minimize inductive self-heating caused by the pulsed magnetic field, small crystals with typical sizes $2 \times 0.5 \times 0.1$~mm$^3$ were cleaved off along the c-direction from the as-grown samples. Data were recorded using a 10~MHz digitizer and 100~kHz alternating current, and analyzed using a custom low-noise digital lock-in technique \cite{Yuan 2009}. Care was taken to ensure that neither the current nor the field pulse caused significant heating. The temperature dependence of the resistivity at zero field was measured with a Lakeshore resistance bridge. Complementary magnetization data $M(T)$ were measured using a Quantum Design SQUID magnetometer.

Figure 1 presents the temperature dependence of the in-plane electrical resistivity $\rho$$_{ab}$$(T)$ for Fe$_{1.11}$Te$_{0.6}$Se$_{0.4}$ at zero field. As reported in the literature \cite{Chen 2008-3, Liu 2009}, Fe$_{1.11}$Te$_{0.6}$Se$_{0.4}$ exhibits a resistivity that increases with decreasing temperature. Nevertheless, it undergoes a relatively sharp superconducting transition at $T_{\rm c}=14\pm 0.3$K. Bulk superconductivity is confirmed by the temperature dependence of the dc magnetic susceptibility, as plotted in the inset of Fig.1.

\begin{figure}
  \includegraphics[width=8cm]{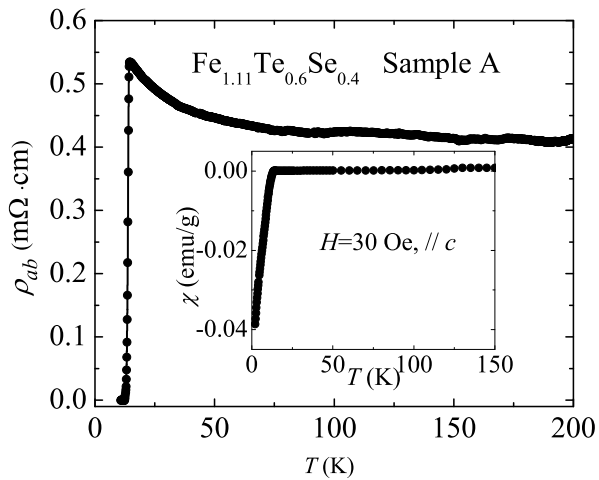}\\
  \caption{(a) Temperature dependence of the in-plane resistivity  $\rho$$_{ab}$$(T)$ at zero field for Fe$_{1.11}$Te$_{0.6}$Se$_{0.4}$ single crystals. The inset shows dc magnetic susceptibility $\chi$(T) measured at 30Oe with ZFC process. Both resistivity and magnetic susceptibility indicate a bulk superconductivity with T$_{c}$ $\approx$ 14K.}
  \label{}
\end{figure}

\noindent

The field dependent electrical resistivity, $\rho(H)$, at various temperatures is shown in Fig. 2(a) and Fig. 2(b) for magnetic fields applied parallel (${\bf H}$ $\parallel$ ${\bf c}$) and perpendicular to (${\bf H}$ $\perp$ ${\bf c}$) the $\emph{c}$-axis, respectively. For consistency, only data collected during the down-sweep of the magnet are shown. The superconducting to normal transition is visible as a sharp rise in $\rho$; inside the superconducting state, an apparent finite $\rho$ is observed for ${\bf H}$ $\parallel$ ${\bf c}$, but not for ${\bf H}$ $\perp$ ${\bf c}$. The former behavior is likely to be due to dissipation associated with thermally-activated flux motion \cite{Tinkham 1975}. Nevertheless, it is obvious that at the same temperature, superconductivity is suppressed by similar values of the magnetic field applied parallel or perpendicular to the c-axis.

\begin{figure}
  \includegraphics[width=8cm]{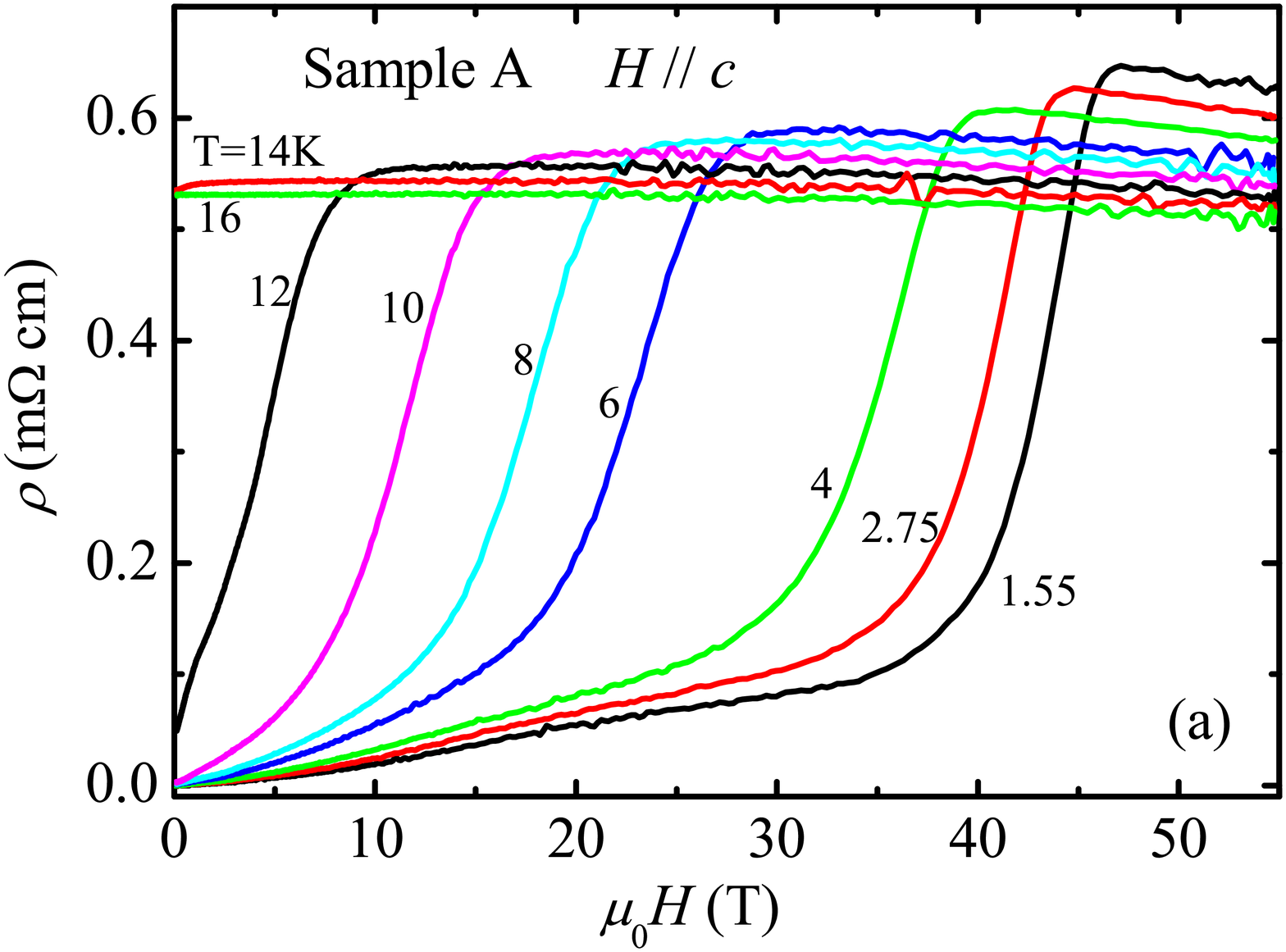}\\
  \includegraphics[width=8cm]{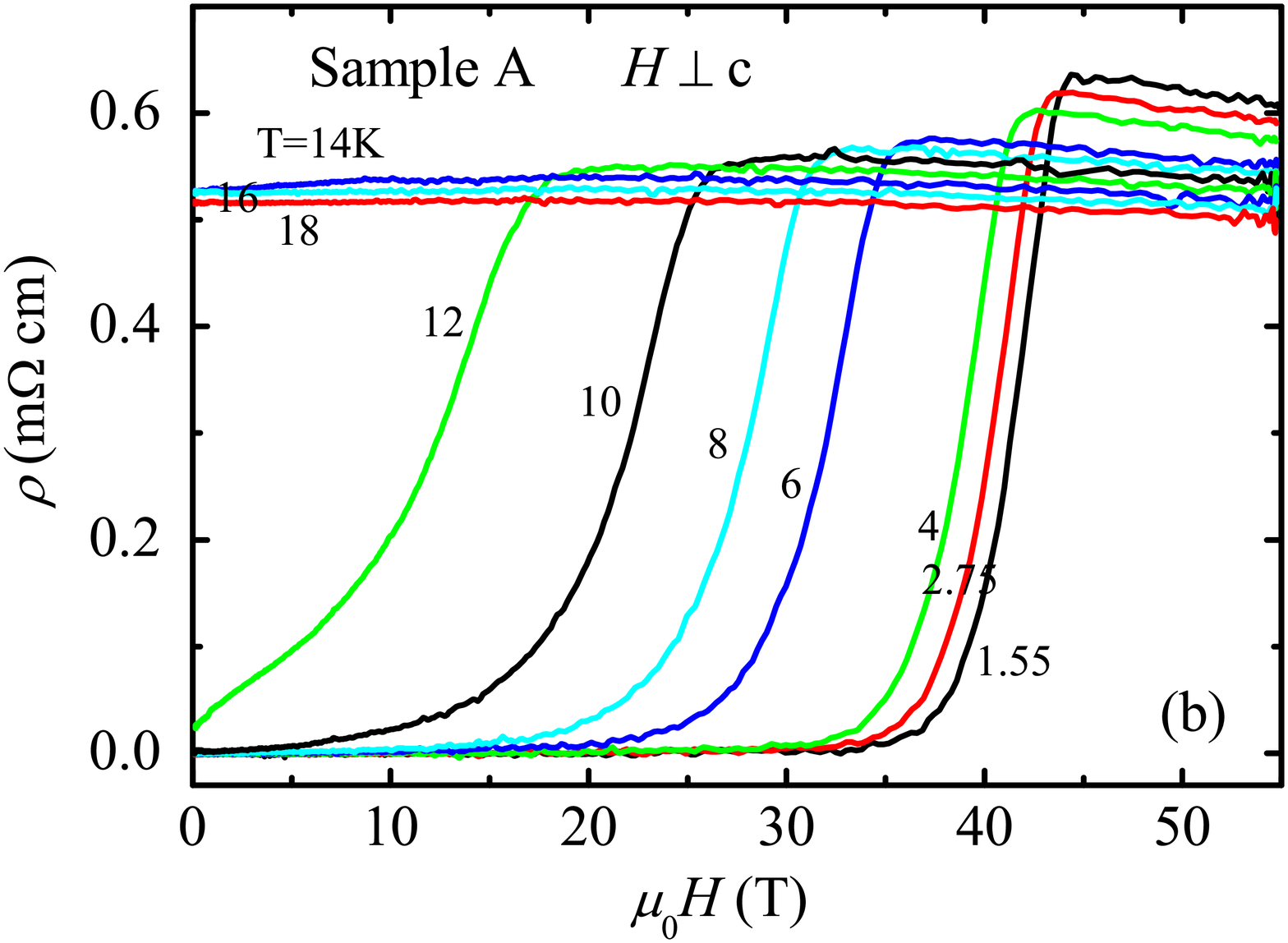}\\
  \caption{The field dependence of the electrical resistivity  $\rho$$_{ab}$(H) at various temperatures for Fe$_{1.11}$Te$_{0.6}$Se$_{0.4}$. (a) $\emph{H}$ $\parallel$ $\emph{c}$; (b) $\emph{H}$ $\perp$ $\emph{c}$.}\label{}
\end{figure}

Figure 3 shows the temperature dependence of the resistivity for various magnetic fields. For a field of 50~T, the superconductivity is suppressed at all temperatures, revealing a normal-state resistivity that increases monotonically with decreasing temperature for both ${\bf H} || {\bf c}$ and ${\bf H} \perp {\bf c}$. This continues the ``insulating'' trend seen at higher temperatures (Fig.~1) which has been attributed to weak charge carrier localization due to the excess Fe \cite{Liu 2009}. However, it should be noted that a weak metal-insulator-like crossover is also observed at cryogenic temperatures in most of the iron pnictides when the superconductivity is suppressed by a large magnetic field \cite{Yuan 2009, Riggs 2008}, suggesting that this behavior might be a more general phenomenon that is not primarily associated with excess Fe in Fe$_{1.11}$Te$_{0.6}$Se$_{0.4}$.

\begin{figure}
  \includegraphics[width=8cm]{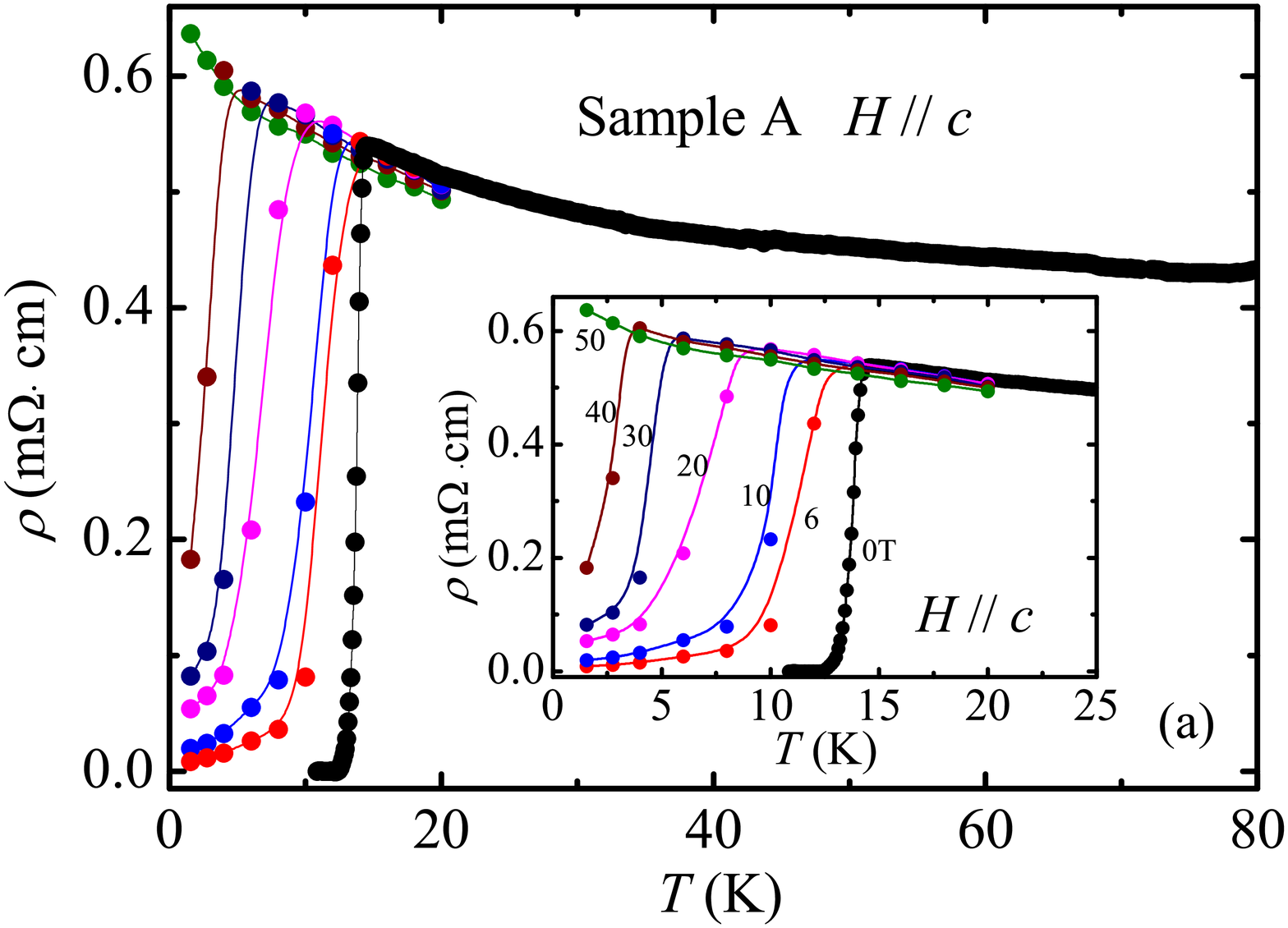}\\
  \includegraphics[width=8cm]{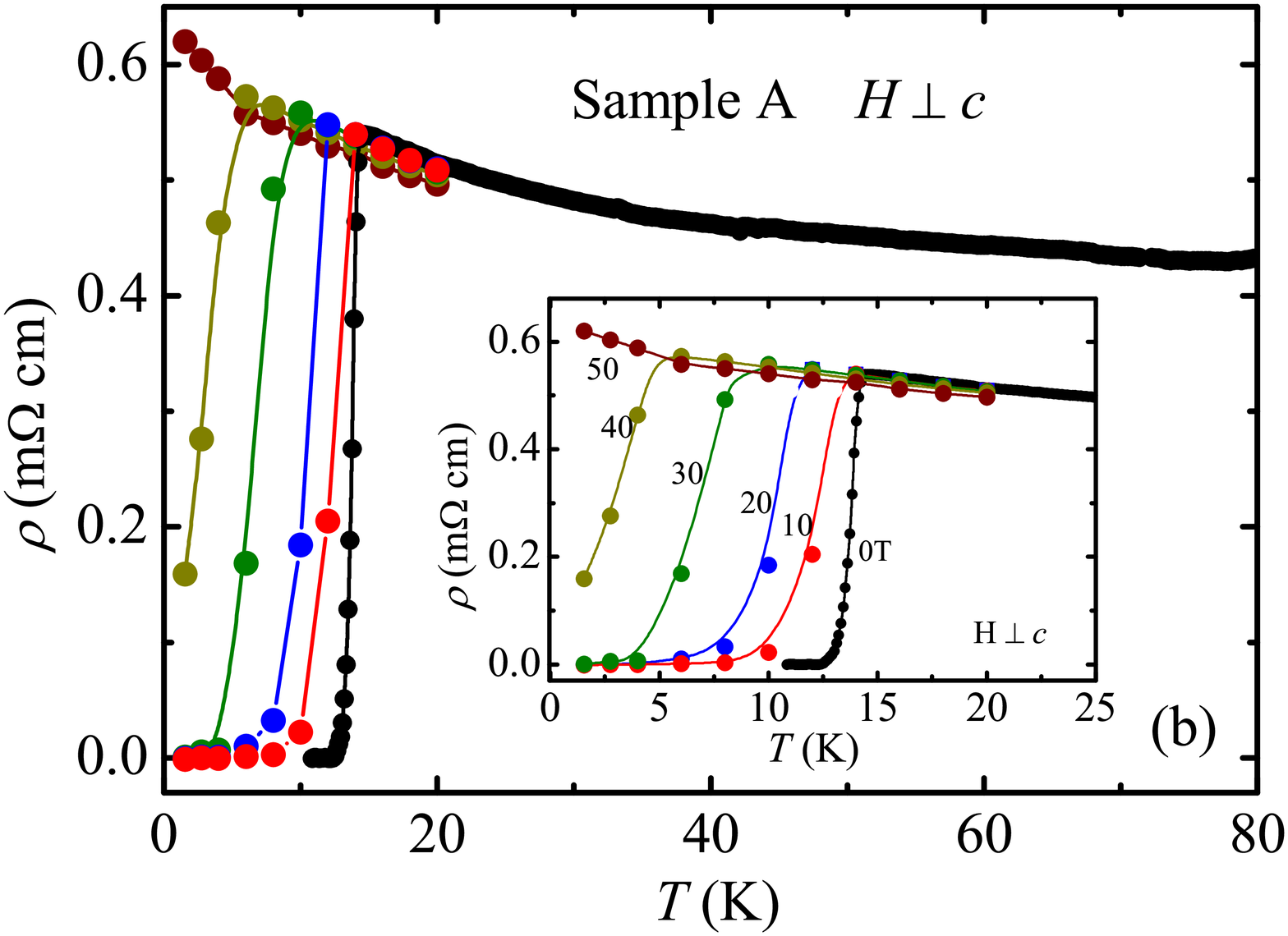}\\
  \caption{The electrical resistivity versus temperature at selected magnetic fields. (a) $\emph{H}$ $\parallel$ $\emph{c}$; (b)$\emph{H}$ $\perp$ $\emph{c}$ axis. The insets plot the superconducting transitions in detail.}\label{}
\end{figure}

The temperature dependence of the upper critical field ($H_{\rm c2}$) of Fe$_{1.11}$Te$_{0.6}$Se$_{0.4}$, determined from the mid-point of the sharp resistive superconducting transitions, as shown in Fig. 2, is plotted in Fig. 4 for magnetic field parallel and perpendicular to the \emph{c}-axis. The two crystals (samples A and B, with $T_{\rm c} =14\pm 0.3$K) exhibit an almost identical behavior of $H_{\rm c2}$, indicating good sample reproducibility. The most remarkable aspect of Fig. 4 is the fact that the upper critical fields of Fe$_{1.11}$Te$_{0.6}$Se$_{0.4}$ for the two field orientations merge together as $T\rightarrow 0$ at $\mu_0H_{\rm c2}\approx 47$~T. This $H_{\rm c2}(0)$ is consistent with the value determined for the polycrystalline sample \cite{Kida 2009}.

\begin{figure}
  \includegraphics[width=8cm]{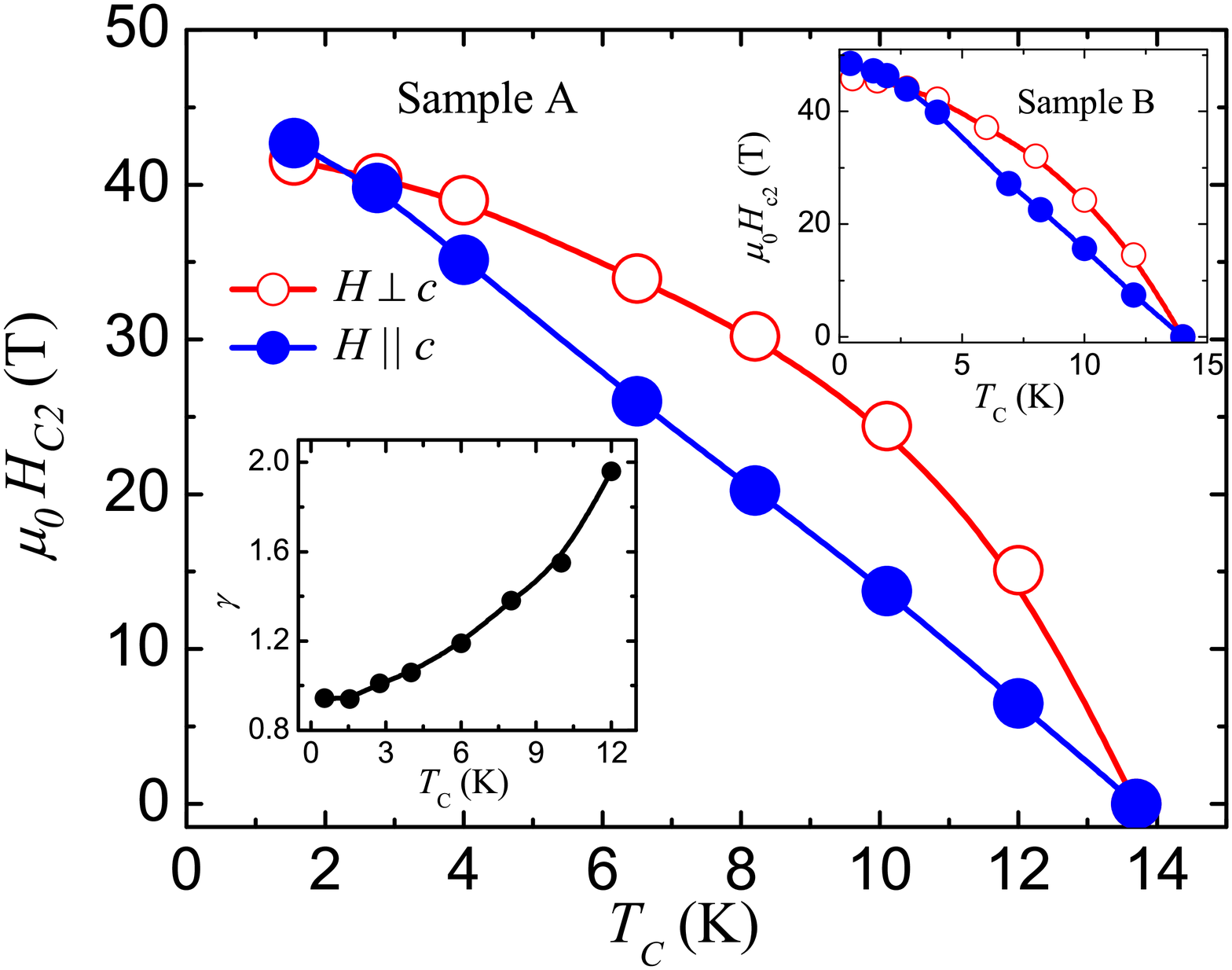}\\
  \caption{Temperature dependence of the upper critical field for sample A (main plot) and sample B (upper inset) where the solid and open symbols represent $\emph{H}$ $\parallel$ $\emph{c}$ and $\emph{H}$ $\perp$ $\emph{c}$, respectively. These data sets indicate good sample reproducibility of $H_{c2}(Tc)$. The lower inset plots the anisotropic coefficient  $\gamma$(=$\emph{H}$$_{c2}$$^{\perp}$/$\emph{H}$$_{c2}$$^{\parallel}$) as a function of temperature for sample B.}\label{}
\end{figure}

The anisotropy coefficient $\gamma(T)$, determined from $\gamma(T)=H_{c2}^{\bot}/H_{c2}^{\parallel}$, decreases monotonically from 2 near $T=T_{\rm c}$ to about 0.95 at $T=0$ (see the lower inset of Fig.4). Similar isotropic behavior of the upper critical field has also been observed in the 122-series of Fe-based superconductors \cite{Yuan 2009, Baily 2009, Altarawneh 2008, Kano 2009}. All these results indicate that nearly isotropic superconductivity might be a general, but very unique feature, of the iron-based superconductors.

The anisotropy of the upper critical field is usually determined by the underlying electronic bandstructure. In the layered cuprates and organic superconductors, the Fermi surfaces are rather two-dimensional \cite{Brooks 2006, Kang 1989}. As a result, there is considerable anisotropy; the upper critical field of these materials is large for in-plane fields, being determined by spin mechanisms such as the Pauli paramagnetic limit, but generally much smaller and restricted by orbital mechanisms for other field orientations \cite{Brooks 2006, Singleton 2002}. However, the experiments in this paper show that this is {\it not} the case for Fe$_{1.11}$Te$_{0.6}$Se$_{0.4}$; its upper critical field $H_{\rm c2}$ at low temperature displays only a very weak dependence on magnetic field orientation (see Fig. 4). By contrast, calculations indicate that the iron pnictides and chalcogenides have much more three-dimensional band structures \cite{Singh 2008, Subedi 2008, Ma 2008, Mazin 2008}. While the layered crystal structure is reflected in the generally cylindrical shapes of the Fermi-surface sections, there is very pronounced dispersion in the $k_z$ direction, leading to strong warping, seen both in the theoretical predictions \cite{Singh 2008, Subedi 2008, Ma 2008, Mazin 2008} and in ARPES data \cite{Vilmercati 2009}; by contrast, there is very little warping in the cuprates \cite{Brooks 2006} and organics \cite{Kang 1989}. We argued in Ref. \cite{Yuan 2009} that the almost isotropic superconductivity observed in (Ba,K)Fe$_2$As$_2$ probably reflects the three-dimensional nature of the Fermi surface. The $\alpha$-Fe(Se,Te) system has the simplest crystal structure of all the Fe-based superconductors, comprising a continuous stack of tetrahedral Fe(Se,Te) layers along the c-axis; consequently, it is expected that the Fermi surface will also be three dimensional in nature, leading naturally to the weak anisotropy of $H_{\rm c2}$ seen here in Fe$_{1.11}$Te$_{0.6}$Se$_{0.4}$. Therefore, the remarkable lack of anisotropy in $H_{\rm c2}$ observed in both the 122- and 11-type iron superconductors has a common origin. In the iron-based superconductors, the coupling between the FeAs or Fe(Te, Se) layers would play an important role and cannot be neglected, which is quite distinct from the cuprates in terms of effective dimensionality.

Although the low-temperature upper critical field is rather isotropic, the initial slope of $H_{\rm c2}$ near $T_{\rm c}$ does show some dependence on the field orientation (Fig. 4); similar behavior in the 122 compounds has been attributed to two-band superconductivity \cite{Baily 2009}. In our resistive critical field data, ${\rm d}H_{\rm c2}/{\rm d}T(T=T_{\rm c}$) is about 8.90~T/K for ${\bf H}\perp {\bf c}$ and 3.82~T/K for ${\bf H} ||{\bf c}$, respectively. These are close to the values observed for Fe$_{1.11}$Te$_{0.7}$Se$_{0.3}$ in dc field measurements \cite{Chen 2008-3, Kida 2009}. Upon cooling down, $H_{\rm c2}(T)$ for ${\bf H}\perp{\bf c}$ starts to bend down, resulting in a significantly lower zero temperature upper critical field compared to typical extrapolation methods. For example, the upper critical field at $T=0$ determined by the Werthamer-Helfand-Hohenberg (WHH) theory \cite{Werthamer 1966} yields a value of about 87~T for ${\bf H}\perp {\bf c}$ (sample A), a much higher value than the actual measured 47~T. It is noted that the multi-band nature of Fe$_{1.11}$Te$_{0.7}$Se$_{0.3}$ may cause a deviation of $H_{\rm c2}(T))$ from WHH theory. From this experimental value of $H_{\rm c2}(0)$, one can calculate the superconducting coherence length of Fe$_{1.11}$Te$_{0.6}$Se$_{0.4}$ as 2.65~nm.

In summary, we have determined the resistive upper critical field of Fe$_{1.11}$Te$_{0.6}$Se$_{0.4}$ single crystals, for fields applied both parallel and perpendicular to the \emph{c}-direction. It is found that the anisotropy of the upper critical field decreases with decreasing temperature, becoming rather isotropic at low temperature ($\mu_0H{\rm_{c2}}(0K) \approx 47$T). Similar behavior was also observed in the 122-type iron pnictides \cite{Yuan 2009, Baily 2009, Altarawneh 2008, Kano 2009}. The nearly isotropic superconductivity shown in these layered compounds is probably attributable to the unique three-dimensional nature of their Fermi-surface topology. This is in great contrast to the cases of high $T_{\rm c}$ cuprates and organic superconductors which possess highly anisotropic upper critical fields due to their quasi-two-dimensional band structure. As mentioned in Ref. \cite{Norman 2008}, our findings of isotropic superconductivity together with a rather high upper critical field suggest that the iron-based superconductors are very promising materials for future applications, in particular if $T_{\rm c}$ could be further enhanced above nitrogen temperature.

This work is supported by the National Science Foundation of China (Grant No.10874146, 10974175, 10934005), the National Basic Research Program of China (Grant No. 2009CB929104), the PCSIRT of the Ministry of Education of China (Contract No.IRT0754), the DOE BES program "Science in 100T" and the NHMFL-UCGP. Work at NHMFL-LANL is performed under the auspices of the National Science Foundation, Department of Energy and State of Florida

\end{document}